\documentstyle[epsfig,prl,aps]{revtex}
\begin{document}
\draft
\twocolumn[\hsize\textwidth\columnwidth\hsize\csname
@twocolumnfalse\endcsname

\title{Universes inside a $\Lambda$ black hole}
\author{I. G. Dymnikova and A. Dobosz}
\address{Institute of Mathematics and Informatics,
University of Warmia and Mazury in Olsztyn,\\
Zolnierska 14, 10-561 Olsztyn, Poland; e-mail: irina@matman.uwm.edu.pl}
\author{M. L. Fil'chenkov}
\address{Institute of Gravitation and Cosmology, Peoples' Friendship 
University of Russia,\\
6 Miklukho-Maklaya Street, Moscow 117198, Russia}
\author{A. Gromov}
\address{St. Petersburg Technical 
University, Polytechnicheskaya 29, 195251 St. Petersburg, Russia}

\maketitle

\begin{abstract}
We address the question of universes inside
a $\Lambda$ black hole which is described by a
spherically symmetric globally regular
solution to the Einstein equations with a variable cosmological
term $\Lambda_{\mu\nu}$, asymptotically $\Lambda g_{\mu\nu}$
as $r\rightarrow 0$ with $\Lambda$ of the scale
of symmetry restoration. 
Global structure of spacetime contains an infinite sequence of black 
and white holes, vacuum regular cores and asymptotically flat universes.
Regular core of a $\Lambda$ white hole models the initial stages
of the Universe evolution. In this model it
starts from a nonsingular nonsimultaneous big bang,
which is followed by a Kasner-type anisotropic expansion.
Creation of a mass occurs mostly at the anisotropic stage of
quick decay of the initial vacuum energy. 
We estimate also the probability of quantum birth 
of baby universes inside a $\Lambda$ black hole due to quantum instability
of the de Sitter vacuum.
\end{abstract}
\vspace{0.2cm}
]

{\bf Introduction-}
The idea of a de Sitter core replacing a black hole singularity  
goes back to the 60-s papers by Sakharov who suggested $p=-\varepsilon$ as 
an equation of state at superhigh densities \cite{sakharov}, and by Gliner 
who interpreted $p=-\varepsilon$ as a vacuum equation of state and suggested 
that it could be a final state in a gravitational collapse \cite{gliner}.

In the 80-s several solutions have been obtained by direct matching 
de Sitter metric inside to Schwarzschild metric outside
of a spacelike junction surface ${\Sigma}_0$
of the Planckian thickness
$\delta\sim{l_{Pl}}\sim{10^{-33}}$cm \cite{match,fg}.
Global structure of spacetime in this case
is shown in Fig.1. 
\begin{figure}
\vspace{-8.0mm}
\begin{center}
\epsfig{file=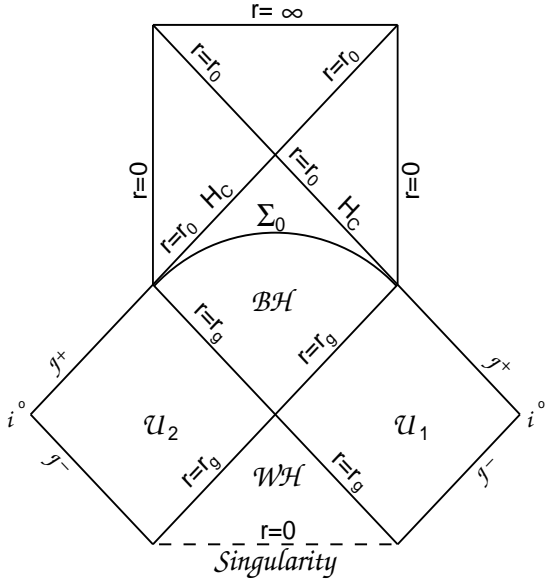,width=7.5cm,height=8.0cm}
\end{center}
\caption{
Penrose-Carter diagram for the case of the direct de Sitter-Schwarzschild
matching.} 
\label{fig1}
\end{figure}  
The idea of a baby Universe inside a black hole 
has been proposed
by Farhi and Guth (FG) in 1987 \cite{fg} as the idea of creation of a universe 
in the laboratory starting from a false vacuum bubble in the Minkowski
space. 
FG studied an expanding spherical de Sitter bubble separated by a thin
wall from the outside region of the Schwarzschild geometry.
The global structure of spacetime in this case (Fig.1) implies
that the expanding bubble must be associated
with an initial spacelike singularity which
clearly represents
a singular initial value configuration. Therefore Farhi and Guth 
concluded that the initial singularity would be
an unavoidable obstacle to creation of 
a universe in the laboratory \cite{fg}. 

In 1988 Poisson and Israel have
analyzed de Sitter-Schwarzschild transition and found
that the spacetime geometry can be self-regulatory and
describable semiclassically down to a few Planckian radii
 by the Einstein equations with a source term 
representing vacuum polarization effects \cite{werner}.
They found also that the Cauchy horizon must exist in this geometry
($H_C$ at the Fig.1). 

In 1989 arising a new universe inside a black hole has been 
considered by Frolov, Markov, and Mukhanov (FMM) \cite{valera1}
in the context of the hypothesis that the curvature is limited by
the Planckian scale and at this scale  
the equation of state becomes $p=-\varepsilon$. 
The difference of FMM from FG approach is that FG assumption of 
existence of a global Cauchy surface may be violated, due to
the existence of the Cauchy 
horizon \cite{werner},
which implies the absence of a global Cauchy surface. 

In 1990 Farhi, Guth and Guven 
studied the model in which the initial bubble is small enough
to be produced without initial singularity \cite{fgg}.
A small bubble classically could not become a universe - instead it would
reach a maximum radius and then contract. FGG
investigated the possibility that quantum effects allow the bubble
to tunnel into the larger bubble of the same mass which
for an external observer disappears
beyond the black hole horizon, whereas on the inside the bubble would
classically evolve to become a new universe \cite{fgg}.
 
Both FG and FMM models are based on matching the Schwarzschild and
de Sitter metrics using thin shell approach
which implies that the whole dynamical evolution from the
equation of state $\varepsilon=p=0$ to $p=-\varepsilon_{Pl}$ 
occurs within a junction layer of the
Planckian thickness $\delta\sim{l_{Pl}}\sim{10^{-33}}$cm.
As a result the matched metrics typically have a jump at the junction
surface.

The general case of a smooth de Sitter-Schwarzschild transition, 
i.e., of a distributed density profile, has been qualitatively addressed
in the paper by Frolov, Markov, and Mukhanov in 1990 \cite{valera2}.
They mentioned two possibilities: or arising a new macroscopic closed universe
either creation of a white hole in a new asymptotically flat universe
which lies in the absolute future with respect to the original
asymptotically flat universe. 

The exact analytic solution describing de Sitter-Schwarzschild transition 
in general case of a distributed density profile, has been 
found by one of us \cite{me 92} in a simple semiclassical model for  
density profile due to vacuum polarization in a spherically symmetric 
gravitational field \cite{me 96}.
This solution belongs to the class of solutions
to the Einstein equations with the source term such that 
$T_r^r=T_t^t;~T_{\theta}^{\theta}=T_{\phi}^{\phi}$.
The stress-energy tensor with such an algebraic structure
describes a spherically symmetric vacuum
invariant under boosts in the radial direction \cite{me 92} 
and represents the extension of the Einstein cosmological term 
$\Lambda g_{\mu\nu}$ to the spherically symmetric $r-$dependent 
cosmological tensor $\Lambda^{\mu}_{\nu}$ \cite{lambda}. 

In the case of de Sitter-Schwarzschild transition it connects in 
a smooth way two vacuum states: de Sitter vacuum 
$T_{\mu\nu}=(8\pi G)^{-1}\Lambda g_{\mu\nu}$ replacing a singularity
at the origin and Minkowski vacuum $T_{\mu\nu}=0$ at infinity.
This corresponds to $r-$dependent cosmological term evolving from
$\Lambda_{\mu\nu}=\Lambda g_{\mu\nu}$ as $r\rightarrow 0$ 
(with $\Lambda$ of the scale of symmetry restoration
in the origin \cite{haifa}) to
$\Lambda_{\mu\nu}=0$ as $r\rightarrow \infty$. 

In the  Schwarzschild coordinates the metric is given by
\begin{equation}
ds^{2}=
\left(1-\frac{R_{g}(r)}{r}\right)
dt^{2}-\frac{dr^{2}}{1-\frac{R_{g}(r)}{r}}-
r^{2}d\Omega^{2}
\label{interval}
\end{equation}
where $d\Omega^{2}$ is the line element on the unit two-sphere.\\
The function $R_{g}(r)$ represents an $r-$dependent gravitational 
radius
\begin{equation}
R_{g}(r)=8\pi G\int\limits_{0}^{r}\rho (x)x^{2}dx=2 G{\cal{M}}(r)
\label{Rg}
\end {equation}
A density profile $\rho(r)=(8\pi G)^{-1}\Lambda_t^t(r)$ 
should be a smooth function providing the proper asymptotic behaviour of 
$R_{g}(r)$: quick vanishing as $r\rightarrow\infty$ to guarantee finiteness of 
a mass
\begin{equation}
$$M=4\pi \int\limits_{0}^{\infty}\rho (x)x^{2}dx < {\infty};
~~~R_{g}(r\rightarrow\infty)=r_{g},$$
\label{mass}
\end{equation}
where $~ r_g=2GM $ and $M$ is the Schwarzschild mass,
and proper asymptotics as $r\rightarrow 0$ \cite{werner,valera2,me 92}
\begin{equation}
$$R_{g}(r\rightarrow 0)=\frac{r^{3}}{r_{0}^{2}}$$
\label{r3}
\end{equation}
where $r_0$ is the de Sitter horizon defined by
\begin{equation}
r_{0}^{2}=\frac{3}{\Lambda}=\frac{3c^{2}}{8\pi G\rho_{0}}
\label{r0}
\end{equation}
Here $\rho_{0}$ is the vacuum density and $\Lambda=\Lambda_t^t(0)$ 
is the value of the cosmological constant at the origin. 

For any density profile satisfying conditions (3)-(4), the metric (1)
describes a globally regular de Sitter-Schwarzschild geometry, asymptotically
Schwarzschild as $r\rightarrow\infty$, and asymptotically de Sitter
as $r\rightarrow 0$ \cite{me 96,me 99}. 
For cosmological term $\Lambda_{\mu\nu}$ 
with the algebraic structure $\Lambda_t^t=\Lambda_r^r;~
\Lambda_{\theta}^{\theta}=\Lambda_{\phi}^{\phi}$,
responsible for this geometry, 
the inflationary equation of state is satisfied by the radial 
pressure $p_r^{\Lambda}=-\Lambda_r^r$, while the tangential pressure 
$p_{\perp}^{\Lambda}=-\Lambda_{\theta}^{\theta}=-\Lambda_{\phi}^{\phi}$ is
calculated from the conservation equation $\Lambda^{\mu\nu}_{;\nu}=0$,
giving the equation of state for cosmological tensor 
$\Lambda_{\mu}^{\nu}$ \cite{lambda} 
\begin{equation}
p_r^{\Lambda}=-\rho^{\Lambda};~ ~
p_{\perp}^{\Lambda}=p_r^{\Lambda}+\frac{r}{2}\frac{dp_{r}^{\Lambda}}{dr}
\label{p22}
\end{equation}

The cosmological term $\Lambda_{\mu\nu}$ corresponds to
a spherically symmetric vacuum 
$T_{\mu\nu}^{\Lambda}(r)=(8\pi G)^{-1} \Lambda_{\mu\nu}(r)$ 
with the variable density and pressure.
It belongs to the Type I in the classification by Hawking and Ellis
\cite{HE}. For any density profile decreasing monotonically  
($d\rho/dr\leq 0$ everywhere) it satisfies 
the weak energy condition $T_{\mu\nu}u^{\mu}u^{\nu} \geq 0$ 
for any timelike vector $u^{\nu}$, which holds if 
$\rho\geq 0,~\rho + p_k \geq 0 ~(k=1,2,3)$ \cite{HE}.
With the above restriction on a density profile $T^{\Lambda}_{\mu\nu}$
satisfies also the dominant energy condition $T^{00}\geq|T^{ab}|$ 
for each $a,b$,
which holds if $\rho\geq 0, ~-\rho\leq p_k\leq \rho$. These 
conditions imply that the local energy density is non-negative
and each component of the pressure never exceeds the energy density.
The strong energy condition  
$(T_{\mu\nu}-g_{\mu\nu}T/2)u^{\mu}u^{\nu}\geq 0$, which for Type I
holds if $\rho+p_k\geq0,~~\rho+\sum{p_k}\geq 0$, is violated
(i.e. a gravitational acceleration changes sign) at the surface of
zero gravity defined by $2\rho+r d\rho/dr =0$.

In the range of masses 
$M\geq M_{crit}\simeq{0.3M_{Pl}\sqrt{\Lambda_{Pl}/\Lambda}}$ 
the de Sitter-Schwarzschild spacetime has two horizons, an event horizon
$r=r_{+}$ and an internal Cauchy horizon $r=r_{-}$, 
and the metric (1)
describes a $\Lambda$ black hole ($\Lambda$BH)
\cite{me 96,lambda}.
Its global structure is presented in Fig.2.
It contains an infinite sequence of black and white holes,
whose singularities are replaced with future and past
regular cores $\cal{RC}$, and asymptotically 
flat universes $\cal {U}$\cite{me 96}. The Penrose-Carter
diagram Fig.2 is plotted in coordinates related to the
photon radial geodesics. The surfaces ${\cal{J}}^{-}$ and
${\cal{J}}^{+}$ represent their past and future infinities. 
The event horizons $r=r_{+}$ and the Cauchy horizons $r=r_{-}$
are formed by the outgoing and ingoing radial photon
geodesics $r_{\pm}$=const.

It is evident from the Penrose-Carter diagram Fig.2 that inside
a $\Lambda$BH there exists an infinite
number of vacuum-dominated asymptotically flat universes 
in the future of $\Lambda$ white holes.
Geodesic structure of de Sitter-Schwarzschild space-time shows
that the possibility of travelling into other universes through a black hole
interior, discussed in the literature for the case of
Reissner-Nordstr\"om and Kerr geometry (\cite{NF} and references therein), 
exists also in the case of a $\Lambda$BH \cite{geod}.  
\begin{figure}
\vspace{-8.0mm}
\begin{center}
\epsfig{file=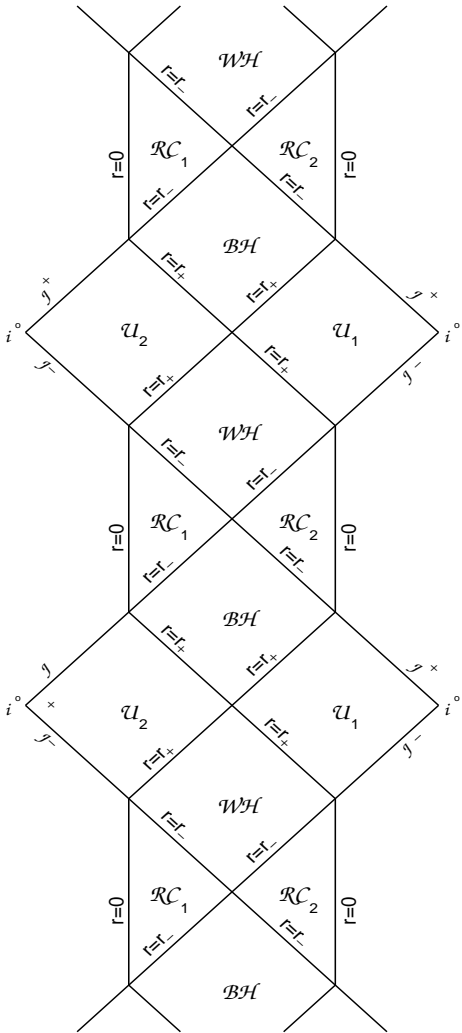,width=9.0cm,height=15.5cm}
\end{center}
\caption{
Penrose-Carter diagram for a $\Lambda$ black hole.} 
\label{fig2}
\end{figure}  
It is widely known that the interiors of white holes can be
described locally as cosmological models (see, e.g., \cite{land}).
In the case of a Schwarzschild white hole it starts from
the spacelike singularity $r=0$ (see Fig.3).
\begin{figure}
\vspace{-8.0mm}
\begin{center}
\epsfig{file=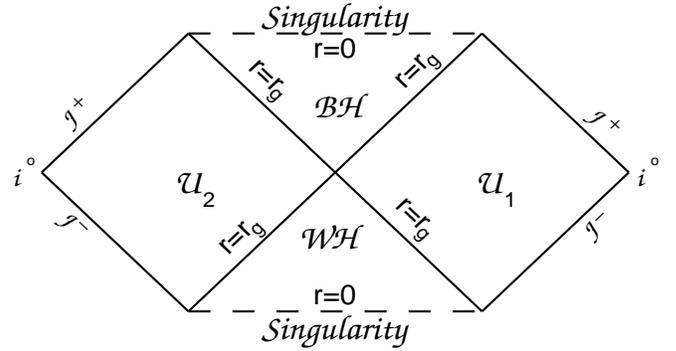,width=9.0cm,height=4.9cm}
\end{center}
\caption{
Penrose-Carter diagram for the Schwarzschild geometry.} 
\label{fig2}
\end{figure}  
Replacing a Schwarzschild singularity with the regular core $\cal{RC}$ 
transforms the spacelike singular surfaces $r=0$, 
both in the future of $\cal{BH}$ and in the past of $\cal{WH}$, 
into the timelike regular surfaces $r=0$ (see Fig.2).
In a sense, this rehabilitates a white hole, whose existence in a singular
version has been forbidden by the cosmic censorship \cite{pen}.

A cosmological model related to a $\Lambda$ white hole corresponds 
to an asymptotically flat vacuum-dominated cosmology 
with the de Sitter origin governed by the time-dependent cosmological 
term $\Lambda_{\mu\nu}$ (segment ${\cal{RC}},{\cal{WH}},
{\cal U}$ in the Fig.2). 
The $\Lambda$WH (more precisely its ${\cal{RC}}$ region) 
models thus the initial stages of a nonsingular 
cosmology with the inflationary origin. 

In this paper we address two questions: First is the 
vacuum-dominated cosmological model with the variable
cosmological term $\Lambda_{\mu\nu}$, related to $\Lambda$WH.
Second is question of baby universes inside a $\Lambda$BH.
The case of direct de Sitter-Schwarzschild matching (Fig.1)
clearly corresponds to arising of a closed or semiclosed world inside a
black hole \cite{valera1,valera2}. We estimate the probability
of this event in the case of global structure Fig.2 as a result 
of quantum instability of the de Sitter vacuum near the surface $r=0$. 
We consider also the case when some admixture of strings or quintessence
is present in the initial fluctuation near $r=0$ 
in the vacuum-dominated system governed by $\Lambda_{\mu\nu}$. 
We show that in such a case a multiple quantum birth of open or flat baby 
universes is possible, which are causally disjoint from each other.
\vskip0.1in
{\bf $\Lambda$WH model for a nonsimultaneous big bang-}

To investigate a $\Lambda$WH together with its past regular core $\cal{RC}$
and future asymptotically flat universe $\cal{U}$, we
introduce the Finkelstein coordinates, related to radial geodesics of 
nonrelativistic test particles at rest at infinity. They are given by
\begin{equation}
c\tau=\pm\,ct\,\pm\,\int\sqrt{\frac{R_{g}(r)}{r}+f(R)}
\frac{dr}{1-\frac{R_{g}(r)}{r}}
\label{podst1}
\end{equation} 
\begin{equation}
R=ct+\int\sqrt{\frac{r}{R_{g}(r)}}\frac{\sqrt{1+f(R)}dr}{1-\frac{R_{g}(r)}{r}}
\label{podst2}
\end{equation}
Here $f(R)$ is an arbitrary function satisfying the condition
$~ 1+f(R)>0$. The lower sign in (\ref{podst1}) is for 
outgoing geodesics 
corresponding to the case of an expansion.

The metric (\ref{interval}) transforms into the Lemaitre metric
\begin{equation}
ds^{2}=c^{2}d\tau^{2}-e^{\lambda(R,\tau)}dR^{2}-r^{2}(R,\tau)d\Omega^{2}
\label{interval1}
\end{equation}
with
\begin{equation}
e^{\lambda(R,\tau)}=\frac{R_{g}(r(R,\tau))}{r(R,\tau)}
\label{lambda}
\end{equation}
Coordinates $R,\tau$ are the Lagrange (comoving) coordinates 
of a test particle, and $r$ 
is its Euler radial coordinate (luminosity distance). In the case of outgoing 
geodesics the $(R,\tau)$ coordinates with the lower sign in (\ref{podst1}), 
map the segment ${\cal{RC}}, {\cal{WH}}, {\cal{U}}$, 
i.e. $\Lambda$WH with its regular  core $\cal{RC}$ and 
its external universe $\cal{U}$.

For the metric (\ref{interval1}) the Einstein equations reduce to \cite{land}
\begin{equation}
8\pi G p_{r}=\frac{1}{r^{2}}\left(e^{-\lambda}r'^{2}-2r\ddot{r}-
\dot{r}^{2}-1\right)
\label{p1}
\end{equation}
\begin{equation}
8\pi G p_{\perp}=\frac{e^{-\lambda}}{r}
\left(r''-\frac{r'\lambda'}{2}\right)-
\frac{\dot{r}\dot{\lambda}}{2r}-\frac{\ddot{\lambda}}{2}-
\frac{\dot{\lambda}^{2}}{4}-\frac{\ddot{r}}{r}
\label{p2}
\end{equation}
\begin{equation}
8\pi G \rho=-\frac{e^{-\lambda}}{r^{2}}
\left(2rr''+r'^{2}-rr'\lambda'\right)+\frac{1}{r^{2}}
\left(r\dot{r}\dot{\lambda}+\dot{r}^{2}+1\right)
\label{ep}
\end{equation}
\begin{equation}
8\pi G T^{r}_{t}=\frac{e^{-\lambda}}{r}
\left(2\dot{r}'-r'\dot{\lambda}\right)=0
\label{T}
\end{equation}
Here the dot denotes differentiation 
with respect to $\tau$ and 
the prime with respect 
to $R$. The component $T_{t}^{r}$ of the stress-energy tensor vanishes
in the comoving reference frame, and Eq.(\ref{T}) is  
integrated giving \cite{tolman}
\begin{equation}
e^{\lambda}=\frac{r'^{2}}{1+f(R)}
\label{lambda1}
\end{equation}                                                             
Putting (\ref{lambda1}) into into (\ref{p1}), we obtain the equation of motion
in the form
\begin{equation}
{\dot r}^2+2r{\ddot r}+\kappa p_r r^2=f(R)
\label{general}
\end{equation}
This cosmological model belongs to the Lemaitre class 
of spherically symmetric models with anisotropic fluid \cite{lem}. 
The dynamics of our model is governed by cosmological tensor
$\Lambda_{\mu\nu}$ which in this case is time-dependent.

Near the surface $r=0$ the metric (\ref{interval1}) transforms into
the FRW form for any $f(R)$. It reads
\begin{equation}
ds^2=c^2d\tau^2-a^2(\tau)(d\chi^2+\sin^2\chi d\Omega^2)
\label{frw}
\end{equation}
with the de Sitter scale factor
$a(\tau)\sim {\cosh(H_0\tau)}$ for $f(R)<0$,
 $a(\tau)\sim {\sinh(H_0\tau)}$ for $f(R)>0$,
$a(\tau)\sim {\exp{(H_0\tau)}}$ for $f(R)=0$,
where $H_0$ is the Hubble parameter corresponding to the
initial value of $\Lambda$.

In this paper we present numerical results for the case of $f(R)=0$
($\Omega=1$).  For numerical integration of the equiation of motion
we adopt the
density profile in the form \cite{me 92}
\begin{equation}
\rho(r)=\rho_{0}exp\left(-\frac{r^{3}}{r_{0}^{2}r_{g}}\right)
\label{rho}
\end{equation}
The characteristic scale of de Sitter-Schwarzschild space-time
is $r_{*}=(r_0^2 r_g)^{1/3}$, and we normalize $r$ to this scale 
introducing the dimensionless variable $\xi$ by $r=r_{*}\xi$. 
The behaviour of pressures in this case  is shown in Fig.4.
\begin{figure}
\vspace{-8.0mm}
\begin{center}
\epsfig{file=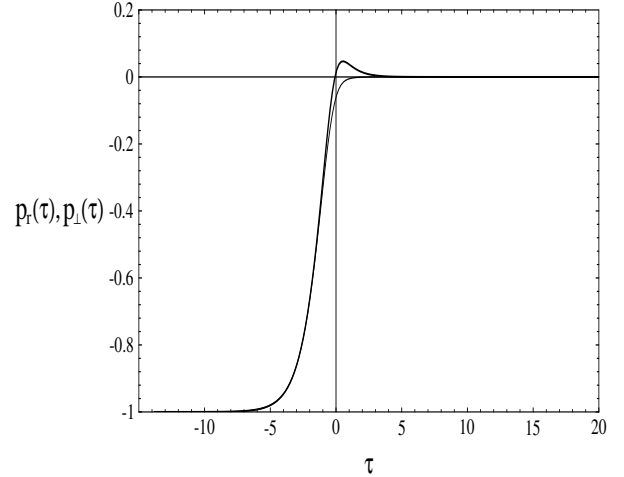,width=8.0cm,height=6.5cm}
\end{center}
\caption{
Radial and tangential pressures $p_r < p_{\perp}$.} 
\label{fig4}
\end{figure}
The equation of motion (\ref{general}) for $f(R)=0$ reduces to 
\begin{equation}
\dot{\xi}^{2}+2\xi\ddot{\xi}-3\xi^{2}e^{-\xi^{3}}= 0
\label{motion}
\end{equation}
It has the first integral
\begin{equation}
\dot{\xi}^{2}=\frac{A-e^{-\xi^{3}}}{\xi}
\label{ksi}
\end{equation}
and the second integral 
\begin{equation}
\tau-\tau_{0}(R)=\int\limits_{\xi_{0}}^{\xi}\sqrt{\frac{x}{A-e^{-x^{3}}}}dx
\label{int}
\end{equation}
Here $\tau_{0}(R)$ is an arbitrary function (constant of integration 
parametrized by $R$) which is called the "bang-time function" \cite{silk}.
For example, in the case of the Tolman-Bondi model for a dust, 
the evolution is described by
$r(R,\tau)=(9GM(R)/2)^{1/3}(\tau-\tau_0(R))^{2/3}$, where $\tau_0(R)$
is an arbitrary function of $R$ representing the big bang singularity
surface for which $r(R,\tau)=0$ \cite{CS98}.

The big bang starts from $\xi_{0}=0$. In our case this 
is the timelike regular surface $r=0$ at the Fig.2. 
Choosing $\xi_{0}=0$ we fix the constant $A=1$ and $\tau_0(R)=-R$. 
In coordinates $(R,\tau)$ the bang starts from the surface $R+c\tau=-\infty$
(see Fig.5).
\begin{figure}
\vspace{-8.0mm}
\begin{center}
\epsfig{file=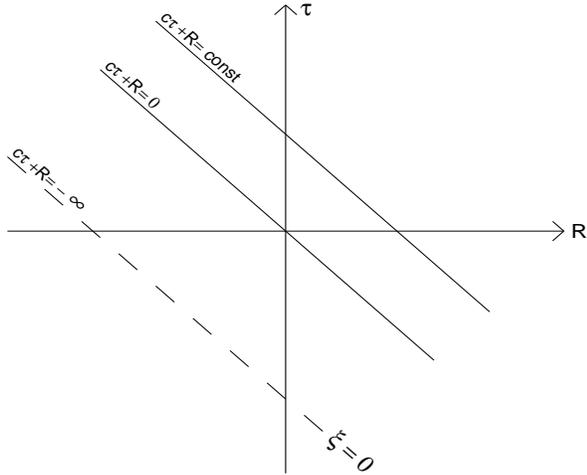,width=8.0cm,height=6.5cm}
\end{center}
\caption{
The Lemaitre metric for a nonsingular white hole. 
Surfaces $r=const$ are plotted for the dimensionless 
radius $\xi$. The surface $\xi=0$ is the big bang surface.}
\label{fig5}
\end{figure}
Different points of the bang surface $\xi=0 ~$ start at different 
moments of synchronous time $\tau$. 
In the limit $\xi\rightarrow 0$ the law of the expansion is
\begin{equation}
\xi=e^{\tau-\tau_{0}(R)}=\dot{\xi}
\label{ksi1}
\end{equation}
This gives
\begin{equation}
e^{\lambda}=\frac{r^{2}}{r_{0}^{2}}\left(\frac{d\tau_{0}(R)}{dR}\right)^{2}
\label{lambda2}
\end{equation} 
and the metric takes the form
\begin{equation}
ds^{2}=c^{2}d\tau^{2}-r^{2}_{0}e^{\frac{2c\tau}{r_{0}}}
\left(dq^{2}+q^{2}d\Omega^{2}\right), 
\label{interval2}
\end{equation}
where the variable $q=e^{\frac{R}{r_{0}}}$ is introduced to transform the 
metric into the FRW form. It describes, with the initial 
conditions $\xi_{0}(R+\tau\rightarrow -\infty)=0$, 
$\dot{\xi}_{0}(R+\tau\rightarrow -\infty)=0$, the nonsingular nonsimultaneous 
de Sitter bang.

In the case of a Schwarzschild WH, a singularity is spacelike (see Fig.1,3),
so there exists the reference frame in which it is simultaneous.
In the case of a $\Lambda$ white hole, a regular surface $r=0$ is timelike, 
and there does not exist any reference frame in which two events occuring
on $r=0$ would be simultaneous.

The first lesson of the $\Lambda$WH model is that the
nonsingular de Sitter big bang must be nonsimultaneous.

The further evolution of the function $\xi$,
velocity $\dot{\xi}$ and acceleration $\ddot{\xi}$
is shown in Fig. 6-8, obtained by numerical integration 
of the equation of motion (\ref{motion}) 
with the initial conditions $\xi_0=10^{-6}$, $\dot{\xi}_0=10^{-6}$. 
\begin{figure}
\vspace{-8.0mm}
\begin{center}
\epsfig{file=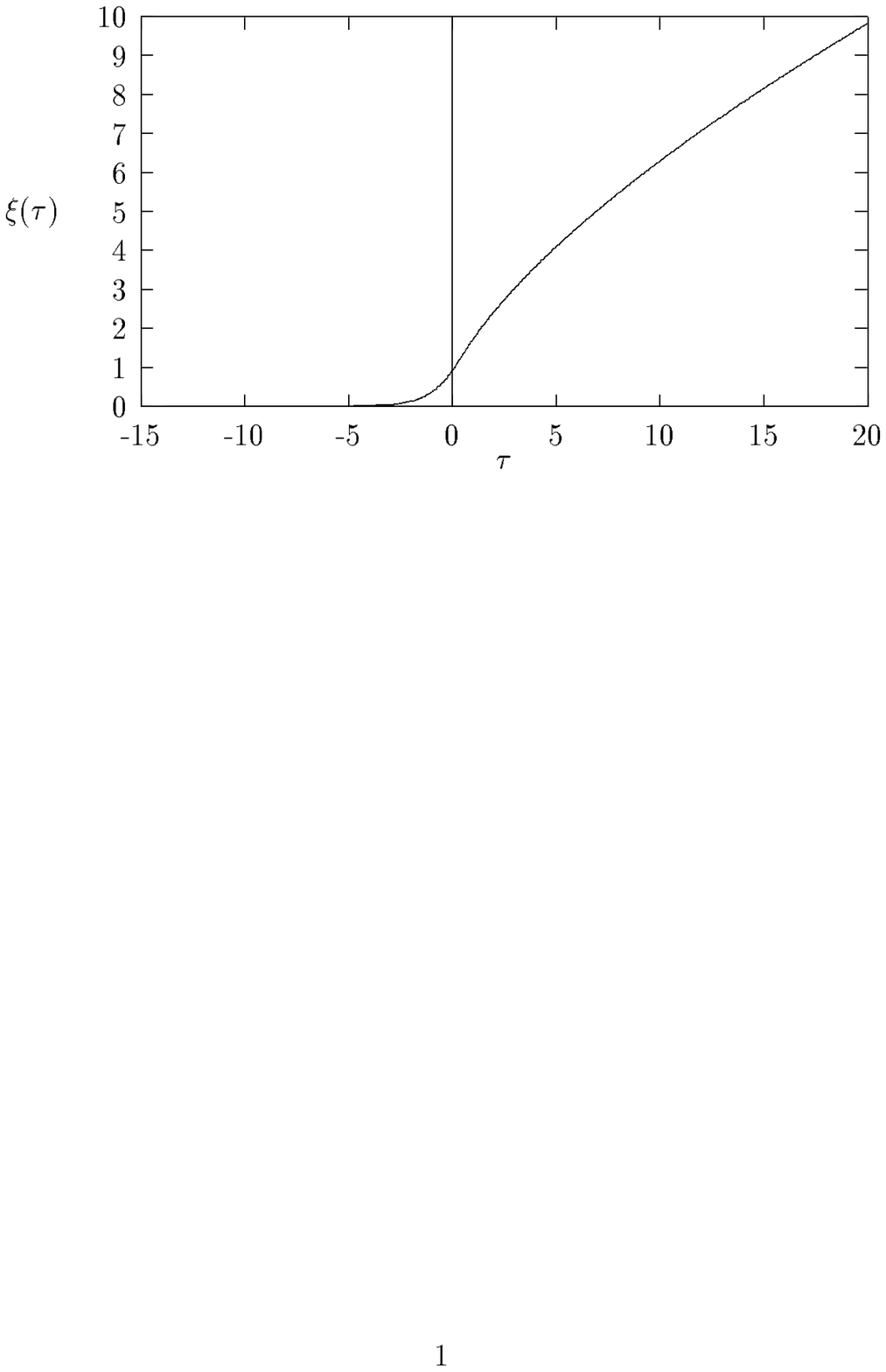,width=8.0cm,height=6.0cm,clip=}
\end{center}
\caption{
The function
$\xi(\tau - \tau_0)$ calculated from the equation of motion
(\ref{motion}) with initial conditions ${\xi}_0= {\dot\xi}_0 = 
10^{-6}$.}
\label{fig6}
\end{figure}
\begin{figure}
\vspace{-8.0mm}
\begin{center}
\epsfig{file=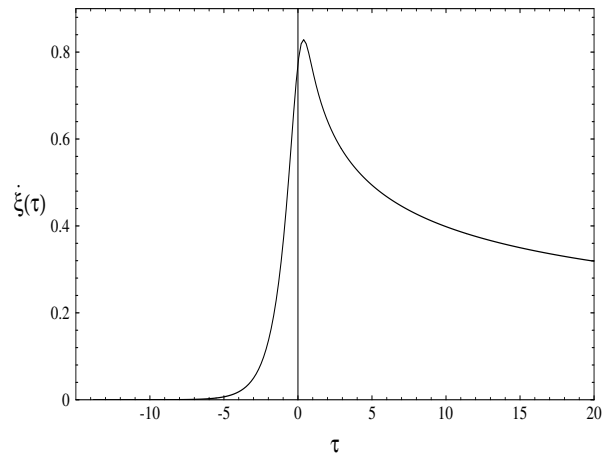,width=8.0cm,height=6.3cm,clip=}
\end{center}
\caption{
The plot of the velocity $\dot\xi(\tau - \tau_0)$ 
for initial conditions ${\xi}_0 = {\dot\xi}_0 = 10^{-6}$.}
\label{fig7}
\end{figure}
\begin{figure}
\vspace{-8.0mm}
\begin{center}
\epsfig{file=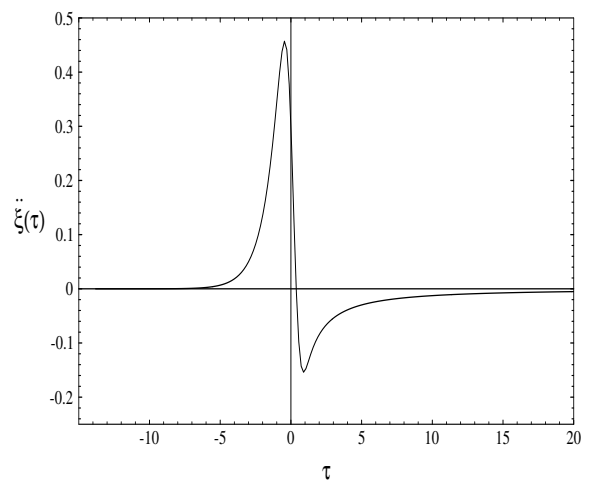,width=7.7cm,height=6.5cm,clip=}
\end{center}
\caption{
The acceleration of the "scale factor" $\xi(\tau-\tau_0)$.}
\label{fig8}
\end{figure}
Numerical integration of the equation (\ref{motion}) shows an
exponential growth of $\xi(\tau-\tau_0)$ at the begining, 
when $p_{\perp}\simeq p_{r}\simeq -\rho$, 
followed by an anisotropic Kasner-type stage when the anisotropic 
pressure leads to an anisotropic expansion.

Qualitatively we can see this approximating the second integral (\ref{int})
in the region $1\ll\xi\ll(r_g/r_0)^{2/3}$ (far beyond a $\Lambda$WH horizon)
by
$$\tau+{R}=\int\limits_{0}^{\xi_{0}(R)}\sqrt{\frac{x}
{1-e^{-x^{3}}}}dx+\int\limits_{\xi_{0}(R)}^{\xi}\sqrt{x}dx$$
It gives
\begin{equation}
\xi=\left(\frac{9}{4}\right)^{\frac{1}{3}}\left(\tau
+\widetilde{\tau}_{0}(R)\right)^
{\frac{2}{3}},
\label{ksi2}
\end{equation}
where 
$\widetilde{\tau}_{0}(R)={R}+
\left(\frac{2}{3}\xi_{0}^{\frac{2}{3}}(R)-F(\xi_{0}(R))\right)$ 
and 
$F(\xi_{0}(R))=\int\limits_{0}^{\xi_{0}(R)}\sqrt{\frac
{x}{1-e^{-x^{3}}}}dx$

Then we get anisotropic Kasner-type metric
$$ds^{2}=c^{2}d\tau^{2}-\left(\frac{9r_{g}}{4}\right)^
{\frac{2}{3}}\left(\tau+\widetilde{\tau}_{0}(R)\right)^{-\frac{2}{3}}
\left(\frac{d\widetilde{\tau}_{0}(R)}{dR}\right)^{2}dR^{2}$$
\begin{equation}
-\left(\frac{9r_{g}}{4}\right)^{\frac{2}{3}}\left(\tau
+\widetilde{\tau}_{0}(R)
\right)^{\frac{2}{3}}d\Omega^{2}
\label{kas}
\end{equation}
with contraction in the radial 
direction and expansion in the tangential direction.

The second lesson of the $\Lambda$WH model is the existence
of the anisotropic Kasner-type stage after inflation.

In our case the Kasner-type stage follows the stage of the nonsingular 
nonsimultaneous big bang from the regular surface $r=0$. 
It looks that this kind of behaviour is generic for cosmological models 
near the origin \cite{BKL} (for recent review see \cite{BGM}). 
Our case differs from a singular case also in that the solution 
is not vacuum in the sense of zero source term in the Einstein 
equations, although it is vacuum in the sense that 
the variable cosmological term 
$\Lambda_{\mu\nu}$ corresponds to a spherically symmetric vacuum 
invariant under boosts in the radial direction \cite{me 92,lambda}.
 
Since the 3-curvature is zero for the case $f(R)=0$, 
the Schwarzschild (ADM) mass ${\cal M}_{ADM}$ coincides with 
the total proper (invariant) mass
${\cal M}_{inv}$ which is the sum of the invariant masses of all particles 
with radial coordinates less than $R$ given by \cite{bondi}
$${\cal M}_{inv}(R)=\int\limits_{0}^{2\pi}d\varphi \int\limits_{0}^{\pi}
d\vartheta \int\limits_{0}^{R}\rho\sqrt{-g}dR=4\pi \int\limits_{0}^{R}
e^{\frac{\lambda}{2}}r^{2}\rho dR$$
Indeed, for a fixed time section, $r$ may be regarded as the 
function of $R$ only \cite{bondi}. Then $e^{\lambda/2}dR=dr$ and
$${\cal M}_{inv}=4\pi \int\limits_{0}^{r}\rho x^{2}dx=
{\cal M}_{ADM}={\cal M}(r)$$
At the beginning ${\cal M}=0$ and $\dot{\cal M}=0$, 
as it follows from the first integral (\ref{ksi}) which gives
$$\dot{\xi}^{2}=\frac{{\cal M}(\xi)}{\xi}$$
The behavior of a mass normalized to the Schwarzschild mass $M$
is shown in the Fig.9.
\begin{figure}
\vspace{-8.0mm}
\begin{center}
\epsfig{file=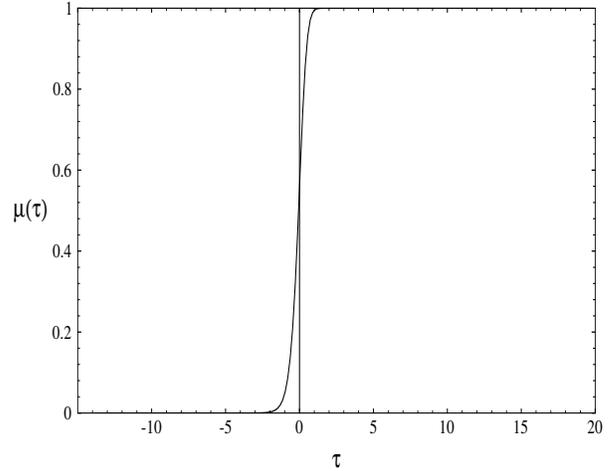,width=8.0cm,height=6.5cm}
\end{center}
\caption{
Plot of the mass function $\mu={\cal M}/M$.}
\label{fig9}
\end{figure}  
During inflationary stage the mass increases as $\xi^{3}$. 
At the next anisotropic stage the mass is 
growing abruptly towards the Schwarzschild mass $M$. 
Since the density is quickly falling at the same time starting 
from the initial value $\rho_{0}=(8\pi G)^{-1} \Lambda$, the 
growth in a mass is connected with the fall of 
$\rho^{\Lambda}=(8\pi G)^{-1}\Lambda_t^t$,
i.e., with the decay of the initial vacuum energy (the growth
of a universe mass by many orders of magnitude in the course
of decay of the de Sitter vacuum was first noticed
in the Ref \cite{us75}).

The third lesson of the $\Lambda$WH model is the quick growth of the mass 
during the Kasner-type anisotropic stage.
\vskip0.1in
{\bf Baby universes inside a $\Lambda$BH-} 
In the case of direct de Sitter-Schwarzschild matching
the global structure of spacetime (Fig.1) corresponds to arising
of a closed or semiclosed world inside a BH \cite{valera1}.
In general case of a distributed density profile (the global structure
of space-time as shown in Fig.2) 
the physical situation  near the timelike surface $r=0$ is similar to that
considered by Farhi and Guth.
This region, which  is the part of the regular core $\cal{RC}$,
differs from that considered in Ref.\cite{fg,fgg}
by an $r$-dependent density profile. Our density profile
(18) is almost constant near $r\rightarrow 0$ and then quickly
falls down to zero. In Fig.10 it is plotted for the case
of a stellar mass black hole with $M=3M_{\odot}$. 
\begin{figure}
\vspace{-8.0mm}
\begin{center}
\epsfig{file=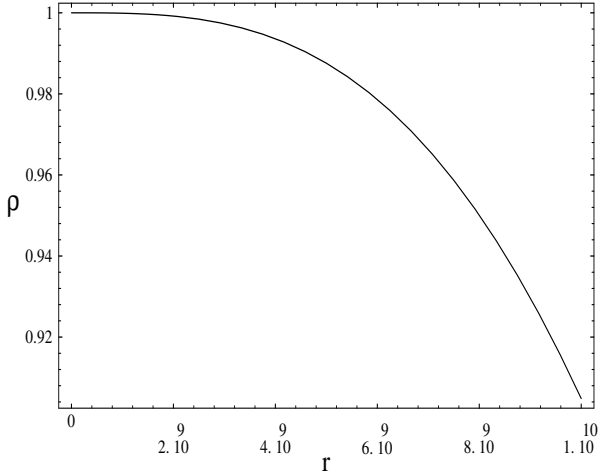,width=8.0cm,height=6.5cm}
\end{center}
\caption{
Density profile (\ref{rho}) for the case of 
three solar masses black hole.}
\label{fig10}
\end{figure}  
We may think of the region near $r=0$ as of a small false vacuum bubble
which can be a seed for a quantum birth of a new universe
in accordance with the main idea of papers \cite{fg,fgg}.
In this case the global structure of space-time is shown in Fig.11, 
which corresponds to arising of a closed
or semiclosed world in one of the $\Lambda$WH structures in the future
of a $\Lambda$BH in the original universe.
\begin{figure}
\vspace{-8.0mm}
\begin{center}
\epsfig{file=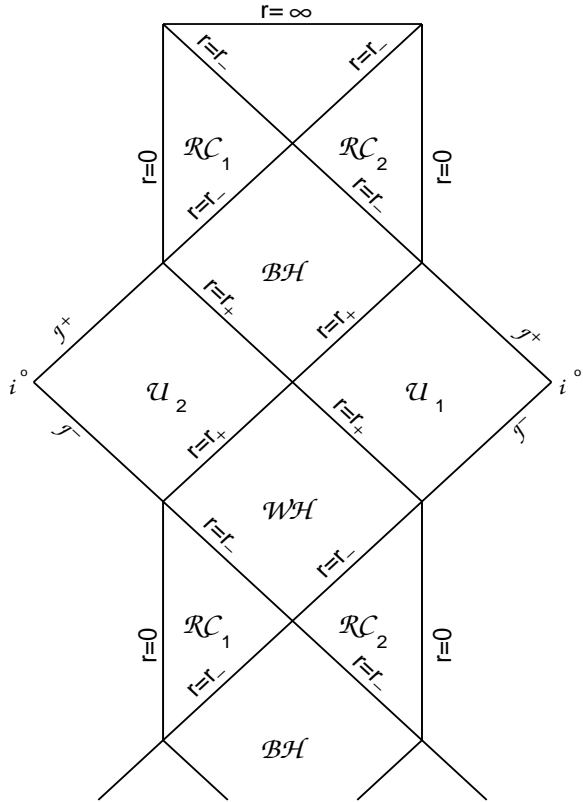,width=8.0cm,height=11cm}
\end{center}
\caption{
The global structure of space-time for the case of a birth of a closed
or semiclosed world inside a $\Lambda$ black hole.}
\label{fig11}
\end{figure} 
On the other hand,
instability of a $\Lambda$ white hole
leads to possibilities other than considered by Farhi and Guth.
Instabilities of Schwarzschild white holes are related
to physical processes (particle creation) near a singularity
(see, e.g., \cite{NF} and references therein).
In the case of a  $\Lambda$ white hole its quantum instability is related
to instability of the de Sitter vacuum near the surface $r=0$.
Instability of the de Sitter vacuum is  well studied,
both with respect to particle creation \cite{bir}
and with respect to the quantum birth of a universe
\cite{us75,gott,lin,albr,vil,vil2,DZS,andrej,olive}.

The possibility of a multiple birth of causally disconnected universes
from the de Sitter background was noticed in 1975 in the Ref. \cite{us75}.
In 1982 such a possibility has been investigated 
by Gott III who considered creation of a universe
as a quantum barrier penetration leading to an open FRW 
cosmology \cite{gott}.
The case of arising of open universes from de Sitter vacuum
is illustrated by Fig.12 \cite{gott}. The events $E$ and $E^{\prime}$
 are creation of causally disconnected universes.
The curved lines are world lines of comoving observers.
At the spacelike surface AB the phase transition occurs
from the inflationary to the radiation dominated stage \cite{us75,gott}.
\begin{figure}
\vspace{-8.0mm}
\begin{center}
\epsfig{file=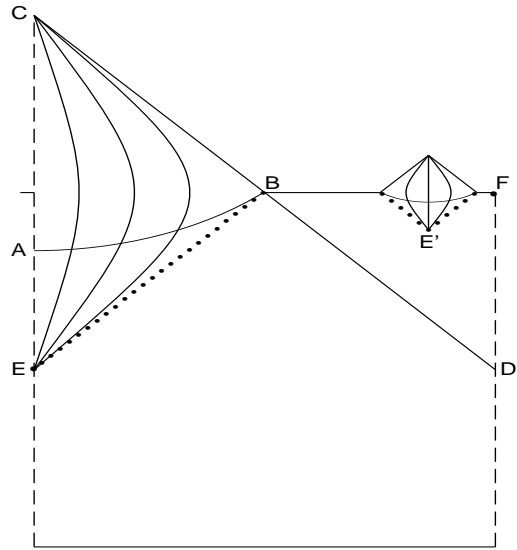,width=7.0cm,height=7.5cm}
\end{center}
\caption{
Penrose-Carter diagram \protect\cite{gott} corresponding to the case 
of a quantum birth of baby universes inside a $\Lambda$BH.}
\label{fig12}
\end{figure} 
In the case of a $\Lambda$BH the region ECB in Fig.12 corresponds to
the region ${\cal{RC}}_1$ in Fig.2, and the region BFD corresponds to
a part of the region ${\cal{RC}}_2$. The regions ${\cal{RC}}_1$
and ${\cal{RC}}_2$ in the de Sitter-Schwarzschild spacetime are entirely
disjoint from each other for the same reason as the regions
${\cal{U}}_1$ and ${\cal{U}}_2$ (they can be connected by space-like curves
only). Birth of open (or flat) baby universes inside a $\Lambda$BH looks 
very similar to the picture shown in Fig.12. The essential difference 
is existence of an infinite number of the regions  ${\cal{RC}}_1$
and ${\cal{RC}}_2$ inside a $\Lambda$BH.

In any case a nucleating spherical bubble  can be
described by a minisuperspace model with a single
degree of freedom, the bubble radius
\cite{vil,vil2}, in our case $a=(r_0^2r_g)^{1/3}\xi$.

The Friedmann equation in the conformal time ($cdt=ad\eta$) reads
\begin{equation}
\biggl(\frac{da}{d\eta}\biggr)^2
=\frac{8\pi G\rho a^4}{3c^2}-ka^2,           
\label{fried}
\end{equation}
where $k=0,\pm 1$. The standard procedure of quantization \cite{vil,vil2}
results in the Wheeler-DeWitt
equation in  the minisuperspace for the wave function of universe \cite{vil}
which reduces to the Schr\"odinger equation 
\begin{equation}
\frac{{\hbar}^2}{2m_{Pl}}\frac{d^2 \psi}{da^2}-[U(a)-E]\psi=0
\label{schrod}
\end{equation}
with $E=0$ and the potential (for the case of $k=1$) 
\begin{equation}
U(a)=\frac{m_{Pl}c^2}{2l_{Pl}^2}\biggl(a^2-\frac{a^4}{r_0^2}\biggr)
\label{pot1}
\end{equation}
With this equation we calculate the probability of a tunnelling 
which describes the quantum growth of an initial
bubble on its way to the classically permitted region $a\geq r_0$,
which corresponds to the case of a closed universe inside a black
hole as in FG and FMM models \cite{fg,fgg,valera1}.

This potential is plotted in Fig.13. It has two zeros, at $a=0$ and $a=r_0$, 
and two extrema: the minimum at $a=0$ and the maximum at $a=r_0\sqrt{2}$.
\begin{figure}
\vspace{-8.0mm}
\begin{center}
\epsfig{file=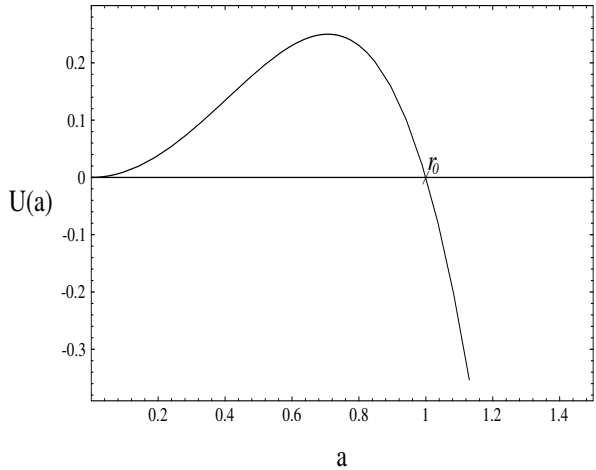,width=8.0cm,height=6.5cm}
\end{center}
\caption{
Plot of the potential Eq.(\ref{pot1}).}
\label{fig13}
\end{figure}  
The WKB coefficient for penetration through the potential barrier reads
\cite{ll}
\begin{equation}
D=\exp{\bigg(-\frac{2}{\hbar}|\int_{a_1}^{a_2}
{\sqrt{2m_{Pl}[E-U(a)]}da}|\biggr)}
\label{d1}
\end{equation}
We get this probability in the frame of the Weeler-DeWitt equation,
applying the general formula for the tunnel effect in quantum mechanics
firstly calculated by Gamow \cite{gamow}.
In the Euclidean approach \cite{HH,vil,vil2} the same result 
for the tunnelling probability  
is obtained \cite{vil2} from the WKB wave function
with the Euclidean Action written in the imaginary time in the
Euclidean domain (under a barrier where the kinetic energy term is negative).
The sign in the exponent in Eq. (\ref{d1})
corresponds to the tunnelling wave function
\cite{vil,vil2}. 

Calculating the integral in (\ref{d1}) we obtain
\begin{equation}
D=\exp{\biggl[-\frac{2}{3}\biggl(\frac{r_0}{l_{Pl}}\biggr)^2\biggr]}
\label{d2}
\end{equation}

For the GUT scale $E_{GUT}\sim{10^{15}}$GeV 
the probability of quantum birth
of a universe is $D=\exp{(-\frac{2}{3} 10^{16})}$. This value of the
penetration factor is in agreement with that calculated in the Ref.
\cite{fgg}. 

To estimate the probability of a quantum birth of an open or flat universe,
we consider the instability of a $\Lambda$WH as evolved from 
a quantum fluctuation near $r=0$ which contains some admixture 
of strings or quintessence \cite{quint} 
with the equation of state  $p=-\rho/3$. 
In this case it is possible to find the nonzero probability 
of tunnelling for any value of $k$ \cite{F}.

In the equation (\ref{fried}) the density evolves with $a$ as
\begin{equation}
\rho=\rho_0\biggl(\frac{a}{r_0}\biggr)^{-3(1+\alpha)},      
\label{eps}
\end{equation}
where
$\alpha$ is a factor in the equation of state
$p=\alpha\rho$.                                          
For the de Sitter vacuum $\alpha=-1$, and $\alpha=-1/3$
for strings or quintessence with $p=-\rho/3$.
When both components are present in the initial fluctuation, 
the density can be written in the form \cite{F}
\begin{equation}
\rho=\rho_0\biggl(B_0+B_2\frac{r_0^2}{a^2}\biggr),         
\label{geneps}
\end{equation}
where $B_0$ and $B_2$ refers to the vacuum and strings (quintessence)
 contributions. The Friedmann equation (\ref{fried}) takes the form
\begin{equation}
\biggl(\frac{da}{d\eta}\biggr)^2=(B_2-k)a^2+\frac{B_0a^4}{r_0^2}
\label{fri}
\end{equation}
and transforms to the Schr\"odinger equation (\ref{schrod})
with the potential
\begin{equation}
U(a)=\frac{m_{pl}c^2}{2l_{pl}^2}\biggl[(k-B_2)a^2-\frac{B_0a^4}{r_0^2}\biggr] 
\label{pot3}
\end{equation}
This potential has two zeros at $a_1=0$ and $a_2=\sqrt{(k-B_2)/B_0}r_0$ 
and two extrema: the minimum
at $a=0$ and the maximum at $a=r_0\sqrt{(k-B_2)/2B_0}$.

The WKB coefficient for penetration through the 
potential barrier (\ref{pot3}) is given by
\begin{equation}
D=exp\left(-\frac{2}{l_{Pl}}|\int\limits_{a_{1}}^{a_{2}}
\sqrt{(k-B_2)a^2-\frac{B_0 a^4}{r_0^2}}da|\right)           
\label{d3}
\end{equation}
We see that the presence of strings or a quintessence with the equation
of state $p=-\rho/3$ in the initial fluctuation  provides
a possibility of quantum creation of flat ($k=0$) and open ($k=-1$)
universe with the probability 
\begin{equation}
D=\exp{\left(-\frac{2}{3}\left(\frac{r_0}{l_{pl}}\right)^2
\frac{\sqrt{(k-B_2)^3}}{B_0}\right)}                         
\label{d4}
\end{equation}

For $r_0\sim 10^{-25}$ cm, $D=\exp{(-\frac{1}{3}\cdot 10^{16})}$ 
for $k=0$, $B_0=2$, $B_2=-1$
which is very close to the value calculated above for the case
of $k=1$ and $B_2=0$, and to that obtained by Farhi, Guth and Guven
\cite{fgg}. 

The probability of a single tunnelling event is very small.
However in the case of a $\Lambda$BH the probability 
of arising of a baby universe in a $\Lambda$BH is the
probability of arising it in one of
the $\Lambda$WH structures inside a $\Lambda$BH.
Since there is an infinite number of $\Lambda$WH in one particular
$\Lambda$BH, this probability is much greater than that
for a single tunnelling event.
\vskip0.1in
{\bf Conclusion -} 
Let us emphasize that an obstacle related to the initial singularity 
does not arise in general case of a distributed profile, since both 
future and past singularities are replaced with the regular surfaces $r=0$.

The results presented above concerning baby universes inside a $\Lambda$ 
black hole are obtained for the case of an eternal black hole. In  case 
a $\Lambda$BH is formed in the course of a gravitational collapse, 
the global structure of the spacetime differs from that shown in Fig.2 
by the presence of a matter. 
The analysis similar to that for the case of direct 
de Sitter-Schwarzschild matching \cite{valera2}, shows
that estimates of probabilities of arising baby universes 
inside a $\Lambda$BH do not change for the case when it arises
in a gravitational collapse.  The probability of a quantum birth of 
baby universes inside a $\Lambda$ black hole is not negligible due to
existence of an infinite number of $\Lambda$WH structures
inside each particular $\Lambda$ black hole.

On the other hand, in the context of creation of a universe
in the laboratory, 
the possibility of influence on the created universe 
is restricted by the presence of the Cauchy horizon in the 
de Sitter-Schwarzschild geometry.

\vskip0.1in

{\bf Acknowledgement}

This work was supported by the Polish Committee for Scientific Research
through the Grant 2.P03D.017.11.

\end{document}